\documentclass[12pt]{article}
\usepackage{sbc-template}
\usepackage{graphicx,url}
\usepackage{listings}
\usepackage[utf8]{inputenc}
\usepackage[T1]{fontenc}
\usepackage[table]{xcolor}
\usepackage{multirow}
\usepackage{float}
\usepackage[para]{footmisc}
\usepackage{parskip}
\usepackage[brazil]{babel}
\usepackage{newfloat}
\usepackage[moderate]{savetrees}

\DeclareFloatingEnvironment[fileext=frm,placement={!ht},name=Code]{listing}
\PassOptionsToPackage{hyphens}{url}

\title{A [in]Segurança dos Sistemas Governamentais Brasileiros: \\ Um Estudo de Caso em Sistemas Web e Redes Abertas}

\author{Marcus Botacin\inst{1} \and André Grégio\inst{1}}

\address{
Universidade Federal do Paraná (UFPR) - 
\texttt{\{mfbotacin, gregio\}@inf.ufpr.br}
}

\begin{document}

\maketitle

\begin{abstract}
Whereas the world relies on computer systems for providing public services, there is a lack of academic work that systematically 
assess the security of government systems. To partially fill this 
gap, we conducted a security evaluation of publicly available 
systems from public institutions. We revisited OWASP top-10 and
identified multiple vulnerabilities in deployed services by scanning
public government networks. Overall, the unprotected services found
have inadequate security level, which must be properly discussed and
addressed.
\end{abstract}

\noindent \textbf{Disclaimer.} 
Nenhum dos sistemas avaliados foi atacado, tampouco houve intenção maliciosa na realização dos testes. O objetivo dos testes realizados foi o  de identificar se os referidos sistemas estavam vulneráveis via acesso Web, com base no documento do OWASP top 10. Tais testes foram verificados previamente a fim de garantir que não haveria violação da integridade, disponibilidade e confidencialidade dos dados dos sistemas avaliados. Todas as vulnerabilidades encontradas foram reportadas aos contatos responsáveis por estes sistemas quando da condução destes (finalizados em maio/2019), bem como posteriormente para o CTIR.gov. Informações sobre os \textit{hostnames} e domínios dos sistemas avaliados foram sanitizadas para preservação de suas identidades.
\section{Introdução}
\label{sec:intro}

Um dos principais agentes da vida 
cotidiana a se beneficiar das plataformas computacionais 
modernas, como a \textit{web} e a Internet, são os órgãos
governamentais, que agora oferecem serviços
\textit{online} aos cidadãos através das plataformas
de governo eletrônico (e-gov)~\cite{Campos:2007:EFE:1352694.1352704}.
Embora eficientes e práticas, a adoção de plataformas
de e-gov também traz novos desafios, como a manutenção 
de níveis de segurança, um requisito fundamental para
sistemas que lidam com dados sensíveis, como as 
informações dos cidadãos. A manutenção dos níveis
de segurança requer um nível alto de maturidade das
práticas de desenvolvimento e de proteção de sistemas,
que devem ser constantemente avaliadas. 

Apesar dos riscos envolvidos na operação de sistemas do 
tipo e-gov já terem sido apontados pela literatura acadêmica
internacional~\cite{Prandini:2011:SCA:2072069.2072119},
ainda são raras as iniciativas voltadas para a avaliação 
da seguranças dos serviços de e-gov no Brasil, uma clara
lacuna de desenvolvimento de impacto não desprezível. 
Em realidade, pode-se observar na prática o impacto
de vazamentos de dados, como os que afetaram os milhares 
de usuários cadastrados no \texttt{Procon}~\cite{procon}. 
Dados do \texttt{CTIR.gov} mostram números não menos
alarmantes sobre a quantidade de ataques recebidos
por órgãos governamentais~\cite{ctir}.

Desta forma, visando dar início a estudos que possam
suprir esta lacuna de conhecimento, este trabalho se 
propõe a conduzir uma avaliação preliminar 
da segurança da infraestrutura dos sistemas computacionais
que suportam os serviços providos por órgãos governamentais
e traçar um panorama das práticas sendo atualmente adotadas. 
Mais especificamente, focamos na avaliação da segurança dos
serviços \texttt{web} e na infraestrutura de rede acessíveis
via Internet, dois pontos
extremamente críticos e reconhecidos pela comunidade como
propensos a apresentar vulnerabilidades. Para a avaliação da 
segurança \texttt{web}, realizamos uma busca \texttt{web} 
por serviços indexados que apresentassem vulnerabilidades 
conhecidas categorizadas pelo projeto \texttt{OWASP top10}~\cite{owasp},
uma ferramenta padrão para a realização deste tipo de
avaliação. Para a avaliação da infraestrutura de rede,
realizamos um escaneamento da redes cujos endereços IPs
são associados a órgão governamentais.

Nossos resultados indicam que o nível de proteção provido
pelos serviços e redes governamentais está abaixo do adequado.
Pudemos identificar a ocorrência das principais vulnerabilidades
\texttt{OWASP} em serviços \texttt{web} colocados em produção
pelos órgãos governamentais. Fomos capazes até mesmo de recuperar
o arquivo de senhas de um dos servidores através de uma vulnerabilidade
do tipo \textit{directory traversal}. Mais ainda, o escaneamento das
redes indicou que muitos servidores ainda se limitam a aceitar tráfego 
não-cifrado, tanto para serviços \textit{web}, quanto para serviços
de e-mail e nomes de domínio. Esperamos que os resultados aqui 
apresentados possam servir de indicadores e guias para a melhoria 
dos níveis de segurança dos serviços providos pelos órgãos 
governamentais. É importante destacar que estas falhas não ocorrem apenas em sistemas computacionais governamentais, mas acreditamos que a investigação desses seja essêncial devido ao alto impacto de eventuais brechas.

Em resumo, este trabalho apresenta as seguintes contribuições:
(i) Uma revisão das principais vulnerabilidades OWASP ilustradas
através de exemplos reais encontrados em sites e serviços governamentais; (ii) Uma análise das principais vulnerabilidades de rede encontradas através de \textit{scans} em larga escala das redes públicas governamentais; e (iii) Uma discussão crítica do estado atual da segurança da infraestrutura
computacional governamental e possíveis ações de melhoria.

\section{Trabalhos Relacionados: Contextualização}
\label{sec:related}

Sistemas de governo eletrônico (e-gov) tem apresentado uma grande
evolução ao longo do tempo~\cite{Campos:2007:EFE:1352694.1352704},
permitindo que governo disponibilizem aos seus cidadãos os mais
diversos tipos de serviço. A possibilidade de agendar eventos
ou mesmo realizar tarefas sem sair de casa torna este tipo
de plataforma muito popular, independente da sua área de 
abrangência. Atualmente, plataformas de e-gov estão disponíveis 
até mesmo em regiões rurais africanas~\cite{Abid:2013:CES:2591888.2591957}.
Como esperado, a tendência de adesão e migração para plataformas de
e-gov também chegou ao Brasil, de modo que o governo federal apresenta
seu próprio portal de serviços eletrônicos disponíveis para os
cidadãos~\cite{egovbr}. 

Ao mesmo tempo em que traz benefícios, plataformas de e-gov também
trazem novos riscos, sobretudo de violações segurança. Eventos como
vazamentos de dados podem violar a privacidade e até mesmo colocar
em risco milhares ou milhões de cidadãos. Apesar disto, pouco tem
se falado da necessidade de se proteger estes tipos de sistemas.
Tipicamente, a segurança de sistemas governamentais tem se voltado
para questões de espionagem~\cite{Verble:2014:NES:2684097.2684101}.
Contudo, neste novo cenário, governos devem também se preocupar contra
ataques oriundos de agentes internos além dos usuais agentes externos,
e esperamos que possamos salientar esta necessidade através deste
trabalho.

Poucos trabalhos encontrados na literatura acadêmica versam sobre
a segurança em sistemas e-gov (e.g., veja um bom exemplo
em~\cite{Prandini:2011:SCA:2072069.2072119}). Dentre estes,
nenhum menciona o cenário Brasileiro, sendo esta uma lacuna
de desenvolvimento que pretendemos suprir através deste
trabalho. No contexto brasileiro, a segurança de alguns 
serviços públicos que afetam uma grande massa de usuários 
começou a ser sistematicamente avaliada recentemente. 
Como resultado significativo, muitas vulnerabilidades 
e más decisões de projeto foram encontradas em aplicações 
bancárias~\cite{Botacin:2019:IBI:3339252.3340103,raju}. Neste 
trabalho, estendemos estas avaliações para os serviços públicos 
disponibilizados via Internet.

\section{Metodologia: Como agem os atacantes?}
\label{sec:meth}

A avaliação foi dividida em duas etapas: (i) inicialmente,
procedemos ao entendimento das principais vulnerabilidades,
suas origens e implicações; (ii) posteriormente, realizamos
uma busca, em maior escala, para identificar a ocorrência
destas na prática.

\noindent \textbf{Entendendo Vulnerabilidades}.
Para o entendimento e caracterização das vulnerabilidades,
adotamos o \textit{framework} \texttt{OWASP top10}~\cite{owasp},
que identifica as vulnerabilidades mais comuns no desenvolvimento
de sistemas \texttt{web}. O \textit{framework} \texttt{OWASP}
tem se mostrado efetivo na identificação e mitigação de vulnerabilidades
em casos reais, sendo adotado tanto academicamente quanto pela 
indústria~\cite{Thai:2019:FWS:3348445.3348456}.

\noindent \textbf{Buscando Vulnerabilidades}.
Para a identificação das vulnerabilidades na prática, buscamos
simular o comportamento de um atacante, que buscaria, primariamente,
por vulnerabilidades conhecidas. Para tanto,
primeiramente realizamos uma busca por \textit{strings} associadas
a vulnerabilidades; por exemplo, buscamos por \texttt{login.php} 
para a identificação de formulários de \textit{login} acessíveis
sem nenhuma proteção. Este tipo de busca é conhecido como
\texttt{google dorks} e tem sido frequentemente utilizado por
atacantes para a exploração em cenários 
reais~\cite{Catakoglu:2017:ALD:3019612.3019796}. Pode-se
encontrar na Internet coleções de \texttt{dorks} para a
identificação de diferentes vulnerabilidades~\cite{dorks}.
Para a identificação das vulnerabilidades de rede,
realizamos um escaneamento das redes cujos IPs são mapeados 
para órgãos governamentais através da ferramenta 
\texttt{shodan}~\cite{shodan}. Esta é uma plataforma de indexação
e busca de dispositivos e serviços expostos na Internet, utilizada
tanto por acadêmicos para a realização de escaneamentos 
direcionados~\cite{Wang:2017:CMP:3078505.3078524}, quanto por
atacantes para a identificação de seus alvos.

\noindent \textbf{Reportando Vulnerabilidades}.
Devemos deixar claro que todos os testes conduzidos tiveram
fins puramente acadêmicos e que nenhuma vulnerabilidade foi 
ativamente explorada de modo a comprometer os sistemas envolvidos. 
Mais ainda, contatamos todos os administradores dos sistemas
analisados de modo a reportar as vulnerabilidades identificadas.
O contato foi realizado com antecedência em relação a 
escrita deste artigo, de modo a garantir tempo hábil para que os 
administradores corrigissem as falhas apontadas. Lamentamos 
que nem todos os envolvidos tenham nos contatados em busca de
maiores informações ou para confirmação das correções.

\section{Vulnerabilidades OWASP: Ocorrências na Prática}
\label{sec:owasp}

Nesta seção, discutimos cada uma das principais 
vulnerabilidades identificadas pelo projeto
\texttt{OWASP} top-10 e exemplificamos a ocorrência destas em
serviços reais de órgãos governamentais brasileiros
expostos à Internet. Destaca-se que múltiplas instâncias
dessas vulnerabilidades foram encontradas (por exemplo, o 
Google apresenta mais de 100 mil e 600 mil resultados para 
páginas contendo os \textit{scripts} \texttt{admin.php} e
\texttt{login.php} em domínios \texttt{.gov.br}), sendo as
apresentadas abaixo selecionadas para serem aqui descritas 
devido ao caráter didático destas.

\noindent \textbf{Ataques de Injeção} são vetores populares de infecção 
de sites e serviços \textit{web}. Eles derivam da ausência ou falha na 
implementação de mecanismos de validação de campos de entrada de dados,
sejam de forma explícita ou implícita, de modo a permitir que um atacante
insira um comando ou trecho de código sob seu controle no fluxo de execução
original do serviço atacado. No passado, os ataques de injeção mais populares
eram as injeções \texttt{SQL}~\cite{Zhang:2018:SIA:3207677.3277958}, na quais 
consultas as base de dados eram abusadas para se obter informações privilegiadas 
(violação de confidencialidade) ou mesmo para executar comandos de administração, 
o que poderia resultar até mesmo na remoção da base de dados (violação de 
integridade). Ao longo do tempo, ataques de injeção também se tornaram populares 
em outras tecnologias \textit{web}, como em \textit{scripts}.

\noindent \textbf{Ataques de injeção do tipo XSS e CSRF} são os tipos mais populares 
de ataques de injeção de código \textit{script} atualmente, dado a popularidade
destas tecnologias em aplicações \textit{web} modernas. Em particular,
a injeção de código \texttt{Javascript} é prevalente dentre os tipos de injeção. Em
ataques XSS, o atacante insere, por exemplo, um \textit{script} em um formulário de 
texto persistente de modo que, quando o site for apresentado para os demais usuários,
o \textit{script} inserido entre em execução~\cite{Gupta:2015:PPM:2742854.2745719}. 
A ocorrência deste tipo de ataque foi identificada na prática, por exemplo, no site 
de uma a prefeitura. O Código~\ref{lst:app:vuln1} ilustra que, quando
da tentativa de acesso ao site original, o \textit{script}
malicioso inserido na página redireciona o usuário para uma página sob controle
do atacante (\texttt{botsqq}). Note que o \textit{script} presente na página
controlada pelo atacante (\texttt{x.php}) recebe o endereço da página original como
parâmetro, permitindo ao atacante contabilizar o número de ``vítimas'' e redirecioná-las
de acordo com a origem, promovendo ataques direcionados.

\begin{lstlisting}[frame=single,label=lst:app:vuln1, caption={Ataque de injeção de
script em página web.} O script inserido redireciona o usuário da página original
para uma página sob controle do atacante.]
Request URL: http://www.xxxxxxxxxxxx.yyy.gov.br/8vLC8QVTzS
Status Code: 302 Found (MOVED) location:
http://botsqq.com/x.php?www.xxxxxxxxxxxx.yyy.gov.br/8vLC8QVTzS
\end{lstlisting}

Além do controle de origem no servidor de destino, descobrimos que o \textit{script} do
atacante também realiza o controle de origem na página atacada, de modo a ocultar a
infecção. Notamos que o \textit{script} não é ativado quando o site é acessado diretamente,
mas apenas quando o visitante é oriundo de outra página. Em termos técnicos, a ativação
do \textit{script} só ocorre quando o cabeçalho \texttt{referer} do protocolo \texttt{HTTP}
está definido, como mostrado no Código~\ref{lst:app:vuln2}.

\begin{lstlisting}[frame=single,label=lst:app:vuln2, caption={Ativação do script malicioso.} O
script só realiza o direcionamento quando o cabeçalho referer do protocolo HTTP esta presente.]
curl -L 'http://www.xxxxxxxxxxxx.yyy.gov.br/8vLC8QVTzS' -H 'Referer: https://www.google.com.br/'
\end{lstlisting}

\begin{figure}[!htpb]
    \centering
    \includegraphics[scale=0.4]{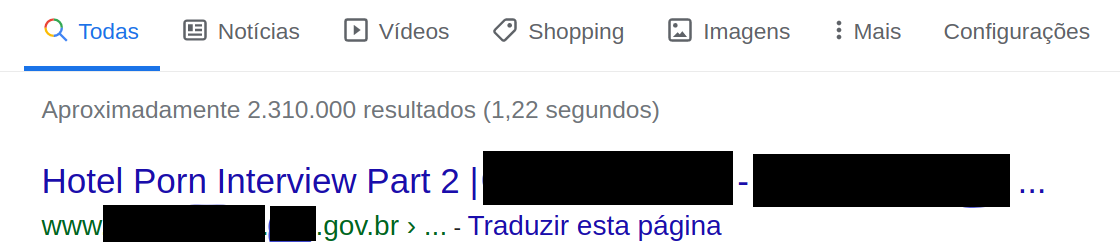}
    \caption{\textbf{Infecção indexada no Google.} Usuários são redirecionados para uma
página de conteúdo adulto.}
    \label{fig:vuln1}
    \vspace{-.5cm}
\end{figure}

O cabeçalho \texttt{referer} é usualmente setado para usuários provenientes de mecanismos de
busca, de modo que infecção pode ser observada em buscadores como o \texttt{Google}, como 
mostrado na Figura~\ref{fig:vuln1}. Neste caso, o atacante redireciona os usuários para uma
página de conteúdo adulto.

\noindent \textbf{Falhas de mecanismos de autenticação e de controle de acesso} são vulnerabilidades
graves que permitem que um usuário acesse recursos não autorizados, o que frequentemente implica na
escalada de privilégios. Estas vulnerabilidades são frequentemente exploradas através de ataques de
injeção de código ou de outras falhas de verificação de entradas. Um tipo de ataque que possibilita
violação de princípios de controle de acesso e autenticação é o \textit{path traversal}, no qual
falhas na verificação das \texttt{URLs} permitem acesso a arquivos internos do servidor. Na prática,
este tipo de ataque foi encontrado no site da prefeitura de uma cidade do Sul do Brasil. O \textit{script}
\texttt{download.php} não trata adequadamente o parâmetro de entrada \texttt{file} e o servidor
\textit{web} não está confinado para rodar em um diretório especial, sendo executado, portanto,
através da raiz (/) do sistema. Deste modo, podemos atravessar a árvore de diretórios até acessar
os arquivos de senhas \texttt{shadow} e \texttt{passwd} 
(\url{https://www.xxxxxxxxx.yy.gov.br/data/download.php?file=../[...]/../etc/passwd}). Deste modo,
um atacante poderia ter acesso a todos os usuários da máquina e decifrar suas senhas em um ataque
de força bruta, ganhando completo acesso ao sistema. Não exibiremos o arquivo \texttt{shadow}, mas
o arquivo \texttt{passwd} é ilustrado no Código~\ref{lst:app:shadow}. 

\newpage

\begin{lstlisting}[frame=single,label=lst:app:shadow, caption={Arquivo de senhas passwd.} Um atacante
poderia se utilizar destas informações para ganhar acesso completo ao sistema.]
SSH:/var/empty/sshd:/sbin/nologin
dovecot:<range>:Dovecot IMAP 
root:x:<range>:root:/root:/bin/bash
cpaneleximfilter:x:<range>::/var/cpanel/userhomes/cpaneleximfilter:/usr/local/cpanel/bin/noshell
postfix:x:<range>::/var/spool/postfix:/sbin/nologin
postgres:x:<range>:PostgreSQL Server:/var/lib/pgsql:/bin/bash
<USERNAME>:x:<range>::/home/<USERNAME>:/bin/bash
\end{lstlisting}

É importante ressaltar que a correção do problema passa pela adotação de práticas de confinamento
(\textit{jail}) do servidor web e da correta filtragem de parâmetros via \textit{script}. Muitas
implementações, contudo, optam por apenas mascarar a vulnerabilidade através de rotinas de ofuscação.
O Código~\ref{lst:app:base} ilustra o caso de site de câmara de vereadores de uma cidade no Sudeste do Brasil, no qual o \textit{script} de \textit{download} aceita parâmetros em \texttt{base64}, tornando
menos evidente o acesso à um diretório. Contudo, a decodificação do mesmo ilustra que o \textit{script}
aceita endereços de arquivo como parâmetro.

\begin{lstlisting}[frame=single,label=lst:app:base, caption={Caminho do arquivo codificado em base64.} Estratégia
de ofuscação não elimina a vulnerabilidade.]
base64 -d http://www.zzzzzzzzzzzzzzzzzz.yy.gov.br/_download.php?file=
aHR0cDovL2NhbWFyYWNhc2NhbGhvcmljby5tZy5nb3YuYnIvdXBsb2Fkcy9kb2N1bWVud
GFjYW8vQXRhLzIwMTcvQXRhLTAxNS5wZGY=
http://zzzzzzzzzzzzzzzzzz.yy.gov.br/uploads/documentacao/Ata/2017/Ata-015.pdf
\end{lstlisting}

\begin{figure}[!htpb]
\begin{minipage}[b]{.45\textwidth}
    \centering
    \includegraphics[width=\columnwidth]{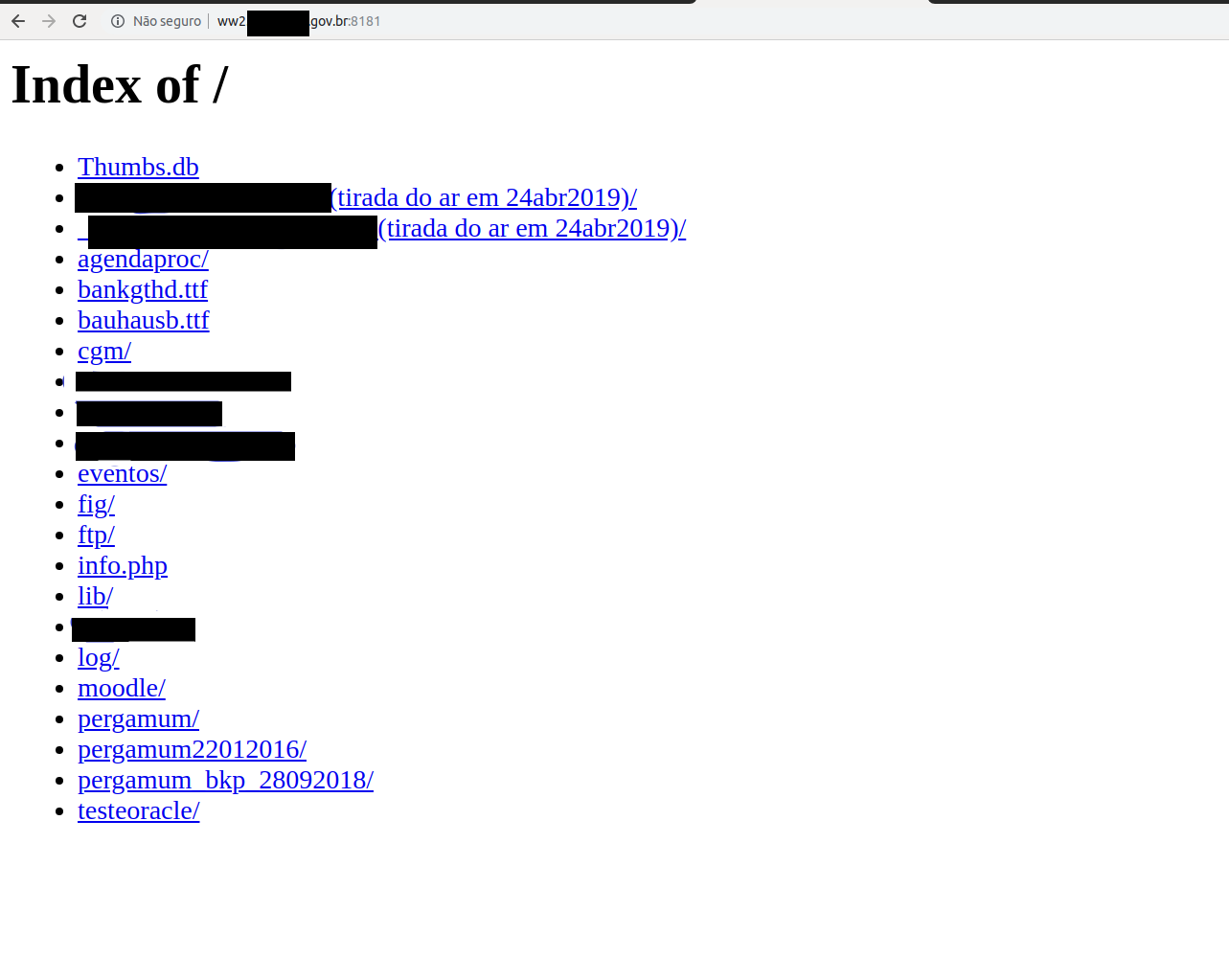}
    \caption{\textbf{FTP de Prefeitura de cidade no Sul do país.} Dados estão
acessíveis sem qualquer mecanismo de autenticação ou controle de acesso.}
    \label{fig:ftp}
\end{minipage}
\hfill
\begin{minipage}[b]{.54\textwidth}
    \centering
    \includegraphics[width=\columnwidth]{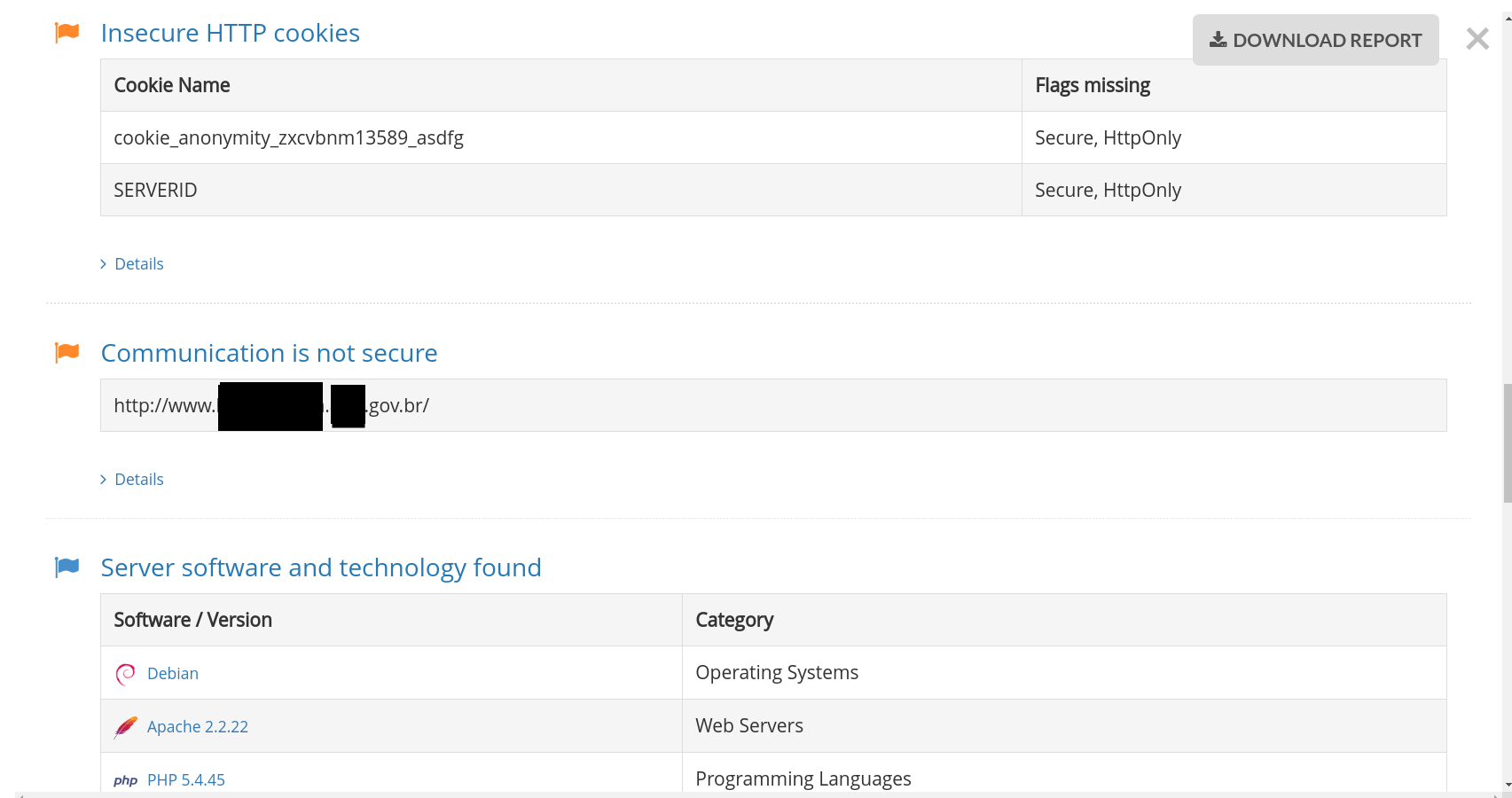}
    \caption{\textbf{Scanner de Vulnerabilidades Web.} Um escaneamento
típico revelaria as múltiplas vulnerabilidades conhecidas presentes no
serviço.}
    \label{fig:vuln2}
\end{minipage}
\vspace{-.5cm}
\end{figure}

\noindent \textbf{Exposição de dados sensíveis} é um resultado frequente
da exploração de vulnerabilidades de implementação ou da ausência de
mecanismos de autenticação e de controle de acesso. Na prática, podemos
observar este caso em servidores de dados do tipo \texttt{FTP} expostos
à Internet, como no caso de outra prefeitura de cidade do Sul do país 
(\url{http://ww2.yyyyyyy.zz.gov.br:8181/}), como mostrado
na Figura~\ref{fig:ftp}.

No caso de órgãos governamentais, é difícil avaliar se os dados foram 
intencionalmente disponibilizados ou não, pois governos seguem políticas 
de dados abertos~\cite{dados1} e de transparência~\cite{transp1}. Desta 
forma, muitos órgãos disponibilizam seus próprios dados em servidores
\texttt{FTP} (e.g., \url{http://xxxxxxxx.xxxxxxx.xx.gov.br/ftp/},
\url{www.yyyyyyy.yy.gov.br/arquivos/ftp}, \url{ftp.zzzzzzzzzzzz.zzz.gov.br}), 
o que é uma boa prática quando estes servidores são bem configurados.
Contudo, a mistura entre arquivos de sistema, dados de séries históricas 
e de licitações correntes mostram a ausência de políticas claras para
a classificação de informações.

\noindent \textbf{Falhas de configuração de mecanismos de segurança} são
um agravante frequente dos incidentes de exploração de vulnerabilidade
\textit{web}, uma vez que a correta configuração destes poderia ter
impedido ou mitigado os efeitos das explorações. Dentre as falhas
mais comuns estão a ausência de cabeçalhos que estabeleçam proteções,
o uso de configurações padrão não otimizadas, a expiração e invalidação
de certificados \texttt{SSL}, entre outros. Na prática, todas essas
práticas foram encontradas no site de uma prefeitura de Estado do Sudeste. Uma varredura de vulnerabilidades
típica~\cite{onlinescan} seria capaz de identificá-las, como
mostrado na Figura~\ref{fig:vuln2}.

De um modo geral, as falhas de configuração em mecanismos de segurança
ocorrem por que estas costumam ser o último fator a ser considerado nos
processos de desenvolvimento~\cite{sdl}. Contudo, em muitas vezes, até
mesmos as etapas de desenvolvimento apresentam falhas significativas.
Não é raro encontrar arquivos de desenvolvimento deixados no servidor 
após o lançamento da aplicação. Fomos capazes de encontrar até mesmos
esquemas de tabelas de banco de dados (\url{http://www.aaaaaaa.aa.gov.br/wp-content/uploads/database.sql}).

\begin{figure}[!htpb]
    \begin{minipage}[b]{.54\textwidth}
    \centering
    \includegraphics[width=\textwidth]{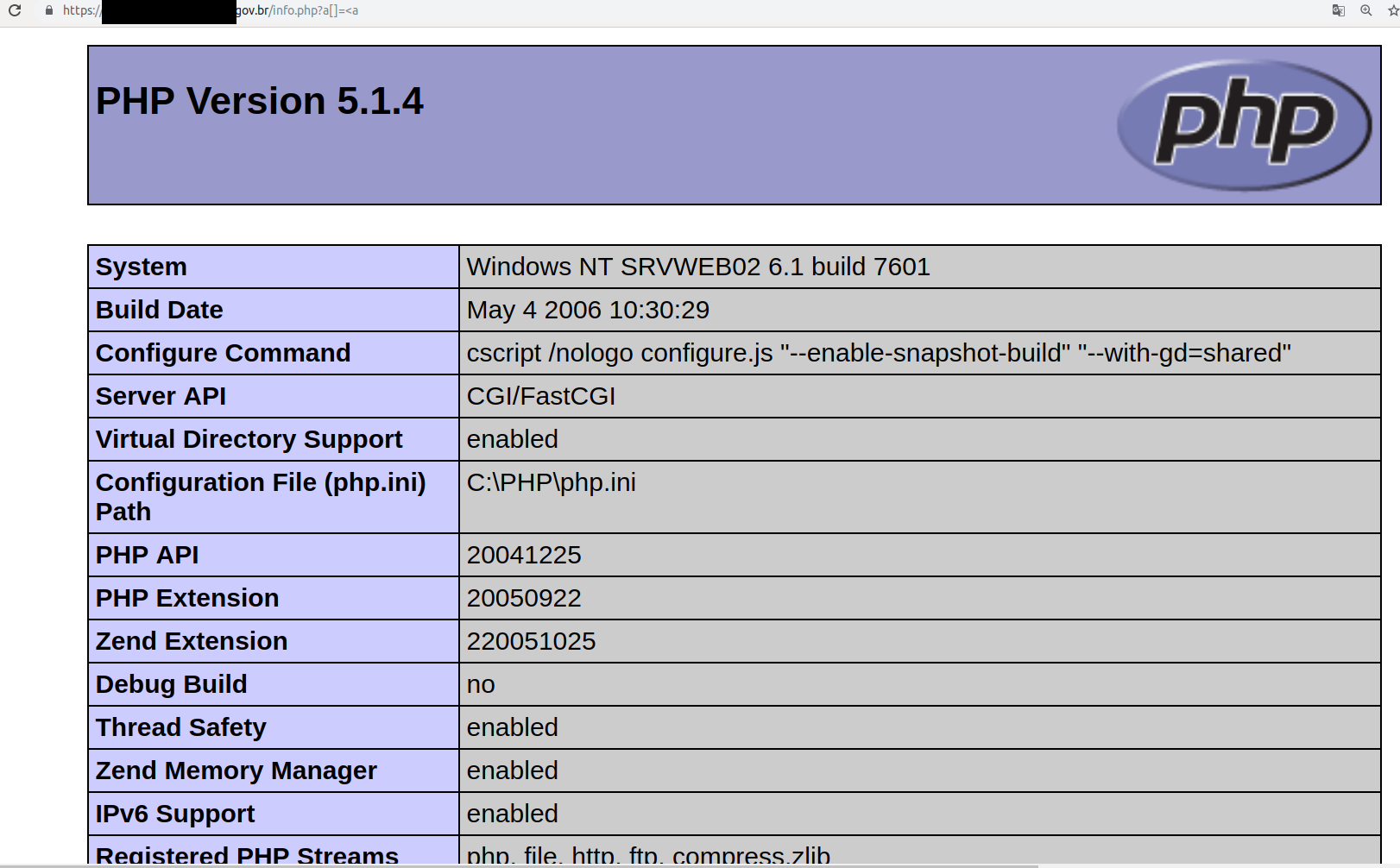}
    \caption{\textbf{Agência de fiscalização governamental.} Sistemas defasados e 
vulneráveis são mantidos \textit{online} expostos a ataques.}
    \label{fig:old_wp}
    \end{minipage}
    \hfill
    \begin{minipage}[b]{.45\textwidth}
    \centering
    \includegraphics[width=\textwidth]{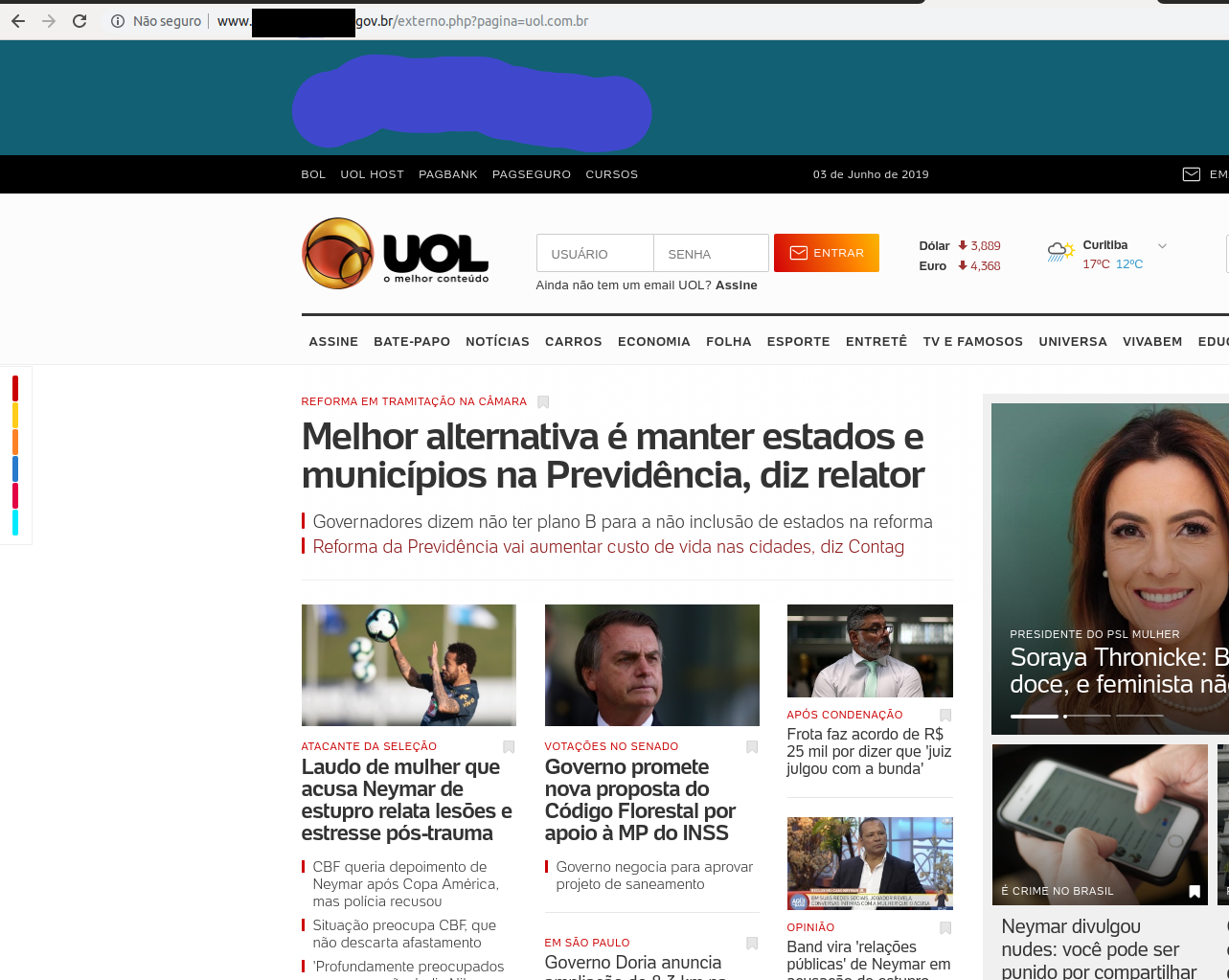}
    \caption{\textbf{Portal de Transparência de Estado do Centro-Oeste.} Parâmetro de redireção na URL
não tratado permite embutir qualquer outro site no corpo do site original.}
    \label{fig:unv_req}
    \end{minipage}
    \vspace{-.5cm}
\end{figure}

\noindent \textbf{Componentes Vulneráveis.}
A atualização de componentes de \textit{software} é um requisito básico para
a manutenção da segurança de qualquer sistema, pois além de novos recursos e 
funcionalidades, atualizações implementam a correção de falhas para impedir
a exploração de vulnerabilidades. Vulnerabilidades em \textit{software} se
popularizam rapidamente via Internet, de modo que atacantes estão sempre
aptos a explorar sistemas que apresentem vulnerabilidades conhecidas por
não terem sido atualizados. Na prática, contudo, ainda podemos encontrar
sistemas defasados expostos na Internet, como mostrado na Figura~\ref{fig:old_wp}.
O site de uma agência de fiscalização de um Governo de Estado do Sudeste, além de vulnerável,
apresenta todas as versões dos serviços em execução por não ter removido
o arquivo de informações do php (\url{https://bbbbbbbbbbbbb.bbbbbb.bb.gov.br/info.php}).
Isto permite que um atacante identifique, com precisão, quais vulnerabilidades
estão presentes naquele servidor.

\noindent \textbf{Requisições e Redirecionamentos Não Validados.}
Serviços e \textit{websites} modernos são construídos de forma modular,
de modo que é comum se encontrar nestes diversos pontos de redirecionamento,
o que permite a comunicação entre componentes e a obtenção de dados de 
diferentes origens. O maior desafio deste tipo de construção é garantir,
através de processos de validação, a origem legítima dos dados. Caso
garantias de validade das requisições não sejam oferecidas, redirecionamentos
podem levar a qualquer outro endereço que não o originalmente previsto
e dados de diferentes fontes podem ser incorporados ao serviço original.
Estas possibilidades são especialmente interessantes à agentes maliciosos,
pois permitem redirecionar o usuário para páginas sob controle do atacante
e/ou injetar código nos sites vulneráveis. Este tipo de vulnerabilidade
é exemplificada, na prática, pelo redirecionamento não tratado no site
do portal da transparência de Estado do Centro-Oeste. Como se pode observar
na requisição \url{http://www.ccccccccccccc.cc.gov.br/externo.php?pagina=site.com.br},
o código \texttt{php} responsável pelo tratamento do redirecionamento não
valida o parâmetro passado, permitindo que qualquer site especificado na
\texttt{URL} seja incorporado ao corpo do site principal, tal como
ilustrado na Figura~\ref{fig:unv_req}.

\noindent \textbf{Log e Monitoração Insuficientes} 
A implementação de mecanismos de monitoração é essencial para o estabelecimento de
processos seguros, sendo que o uso de \textit{logs} pode se dar de forma proativa 
ou responsiva. No primeiro caso, pode-se detectar varreduras, ataques de força 
bruta, e similares, o que permite a identificação do componente sob ataque. No
último, pode-se identificar quais sistemas foram comprometidos e/ou que senhas
foram vazadas, por exemplo, permitindo a resposta ao incidente. A verificação
externa da abrangência e correta implementação dos mecanismos de \textit{log} 
é extremamente dificultada, devendo, portanto, ser realizada prioritariamente de
modo interno ao sistema avaliado. Podemos, contudo, realizar algumas inferências
sob o estado atual dos sistemas governamentais de resposta a incidentes como
um todo. Apesar de estarmos constantemente realizando varreduras nas redes
destes órgãos e seguidas tentativas de conexão a diversos servidores, não
recebemos nenhuma notificação de nenhum destes órgãos, o que pode indicar
que o nível de monitoração não esteja adequado. Nossa hipótese é reforçada
pelo fato de que recebemos notificação da RNP/CSIRT local por sermos origem deste
tipo de ação na rede, o que indica que o efetivo monitoramento das redes
seria capaz de denunciar a ação de um atacante. Mais alarmante ainda,
não recebemos nenhuma notificação mesmo quando acessamos o arquivo de
senhas \texttt{shadow} de alguns servidores, o o que pode indicar que 
o acesso à estes arquivos não eram monitorados a despeito da importância
destes.

\section{Uma Visão Geral dos Serviços de Rede}
\label{sec:overv}

\begin{figure}[!htpb]
    \centering
    \includegraphics[width=.65\textwidth]{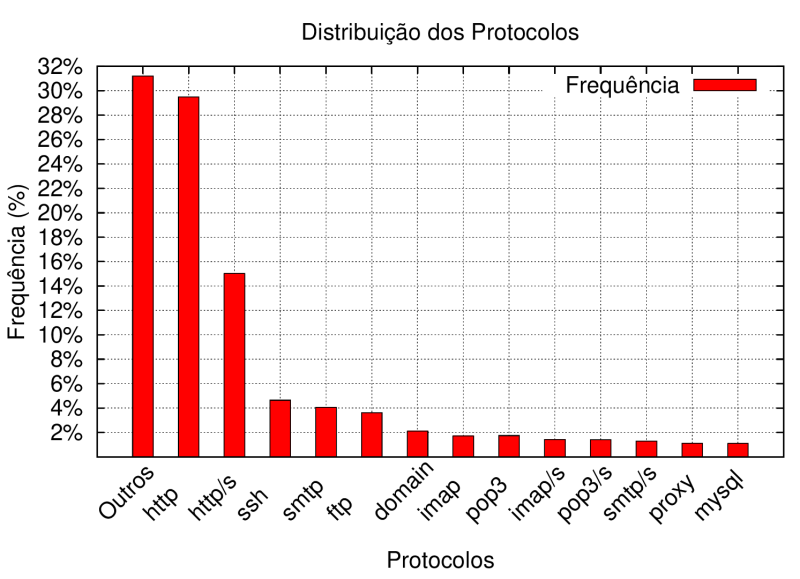}
    \caption{\textbf{Serviços e protocolos identificados.} Muitos
servidores ainda estão configurados de forma insegura, permitindo
apenas tráfego não-cifrado (HTTP na porta 80) ou o encaminhamento
de e-mails de qualquer origem (SMTP na porta 25).}
    \label{fig:prot}
    \vspace{-.5cm}
\end{figure}

Para prover uma visão geral das práticas de segurança adotadas pelos 
órgãos governamentais, estendemos nossa avaliação para incluir o
\textit{deploy} e a configuração de serviços de rede, além das
aplicações \textit{web}. Realizamos uma varredura dos servidores
expostos à Internet através da solução \texttt{shodan} e mapeamos 
os principais serviços disponibilizados e as configurações utilizadas 
por estes. Devido ao grande número de \textit{hosts} conectados à
Internet brasileira, limitamos nossas analises aos 1617 domínios
presentes na lista de domínios \texttt{.br} do portal de dados
abertos~\cite{listadom}. A Figura~\ref{fig:prot} sumariza os serviços e protocolos identificados.

De um modo geral, a maioria dos \textit{hosts} expostos à Internet implementa
servidores \textit{web}, o que é esperado, dado a necessidade de prover
informações a população. Contudo, esperava-se que a maioria destes 
servidores implementassem mecanismos de comunicação segura \texttt{HTTPs},
o que não foi observado na prática. Nota-se que o número de servidores
aceitando conexões do tipo \texttt{HTTP} é muito superior ao número de
servidores aceitando \texttt{HTTPs}, de modo que, mesmo se desconsiderarmos
os servidores que implementam apenas a redireção de tráfego \texttt{HTTP}
para \texttt{HTTPs} na porta 80, ainda temos um número expressivo de servidores
que aceitam apenas conexões não criptografadas. Ressaltamos que, de acordo
com os padrões atuais~\cite{https1,https2}, mesmo informações disponíveis
publicamente e do tipo somente-leitura são consideradas sensíveis, pois a 
simples informação de qual usuário acessou qual tipo de dado pode revelar 
informações significativas sobre este. Infelizmente, os erros de configuração
não se limitam apenas a falta de suporte ao tráfego cifrado, mas estão
pŕesentes também quando este suporte é implementado. Dentre os servidores 
configurados para aceitar conexões criptografadas, 10\% tinham os certificados
\texttt{SSL} vencidos na data de escaneamento (Outubro/2019).

\vspace{-.3cm}
Além de servidores de páginas \textit{web}, outros serviços também são afetados
por má práticas de configuração. Dentre os servidores disponibilizando serviços
de \texttt{e-mail}, 50.7\% aceitam conexões externas não-autenticadas na porta 25.
Desde 2012, recomenda-se o bloqueio da porta 25 para o envio de emails por parte dos clientes para se evitar a propagação de mensagens \texttt{spam}~\cite{spam}. Apesar disso, pudemos nos conectar normalmente a estes servidores.
De modo similar, dentre os servidores provendo serviços
de nomes de domínios (\texttt{DNS}), 47\% respondem a requisições externas, o que
pode propiciar ataques do tipo amplificação~\cite{amplif}. Finalmente, 90.5\% dos
servidores de acesso remoto via \texttt{SSH} estavam configurados na porta padrão
e aceitando autenticação por senha e não por chave pública, sugerindo que estes
estariam sendo executados com as configurações padrão.

\section{Discussão}
\label{sec:disc}

\noindent \textbf{Adotar as melhores práticas de desenvolvimento} de
sistemas é essencial para torná-los mais robusto e menos suscetíveis
a falhas e a apresentar vulnerabilidades. Deve-se, por exemplo, verificar 
e validar todas as entradas providas, uma prática não exaustiva nos sistemas
avaliados. Além disso, deve-se utilizar mecanismos de segurança explícitos,
como recursos de controle de acesso e autenticação para acesso a arquivos, 
e não mecanismos de segurança por obscuridade, como a codificação de 
\texttt{URLs}, como observado em diversos casos.

\noindent \textbf{Adotar as melhores práticas de \textit{deployment}}
de sistemas é essencial para não tornar serviços vulneráveis. Deve-se
evitar usar configurações padrão, não otimizadas para o caso de uso
específico. Deve-se, ainda, remover componentes não utilizados, diminuindo,
assim, a superfície de ataque.

\noindent \textbf{Realizar testes de segurança periódicos} é fundamental
para a identificação tanto de novas vulnerabilidades oriundas de novas
descobertas quando da introdução de falhas de configuração oriundas da
atualização de serviços. A execução deste tipo de deste poderia ter
identificado as vulnerabilidades apontadas neste trabalho.

\noindent \textbf{Antecipar-se aos novos vetores de ataque} 
é essencial para manter a segurança dos sistemas governamentais 
a longo prazo. Enquanto os sistemas atuais se mostram frágeis 
mesmo em relação à ameaças conhecidas, novos ataques vem sendo 
constantemente desenvolvidos. Atualmente, ataques do tipo 
\texttt{XSS}, por exemplo, tem evoluído para ataques do tipo
\textit{data-only}~\cite{Lekies:2017:CAW:3133956.3134091},
sendo ainda mais difíceis de serem prevenidos, detectados e 
eliminados.

\noindent \textbf{Avaliar a maturidade das práticas de segurança 
nas diferentes esferas governamentais} é essencial para um melhor 
entendimento de quais pontos precisam ser fortalecidos e quais 
práticas devem ser prioritariamente adotadas. Empiricamente, 
dado o número de vulnerabilidades encontradas em nosso estudo, 
acreditamos que as esferas municipais tem serviços mais vulneráveis
do que as esferas estaduais e federal. Um hipótese plausível para 
suportar este fato seria os diferentes patamares de investimento 
em cada uma destas esferas. A verificação desta percepção e hipótese
é deixada como trabalho futuro.

\noindent \textbf{Limitações.} Este trabalho se propõe a apresentar um panorama da segurança dos serviços de rede, portanto, particularidades e vulnerabilidades
outras que as presentes no \textit{framework} \texttt{OWASP} não são reportadas,
ainda que estas possam ter impactos significativos na segurança dos sistemas.
Ademais, por este trabalho ser suportado pelos resultados providos pelas 
ferramentas de escaneamento, a identificação de \textit{hosts} também
é limitada pela capacidade destas.

\noindent \textbf{Trabalhos Futuros.} Acreditamos que a avaliação
contínua dos níveis de segurança é essencial para a melhoria das
condições operacionais dos sistemas por prover \textit{feedback}
sobre a eficácia das configurações implementadas. Desta forma,
para o futuro, planejamos repetir estas análises periodicamente
de modo a acompanhar a evolução dos sistemas. Desejavelmente,
o escopo das avaliações será ampliado para a identificação de
outros erros de configurações e falhas.

\section{Conclusão}
\label{sec:conc}

Neste trabalho, avaliamos a segurança dos serviços e da infraestrutura
das redes governamentais expostas na Internet. Revisitamos as principais
vulnerabilidades OWASP e identificamos a ocorrência destas em aplicações
implementadas por diferentes órgãos, o que indica que
muitos destes ainda são incapazes de se proteger mesmo contra ataques e 
vulnerabilidades conhecidas. Realizamos também um escaneamento
das redes e identificamos que grande parte dos servidores ainda se limita
a aceitar dados não-criptografados, seja em serviços \textit{web}, seja
em serviços de e-mail e nomes de domínio. Diante destas descobertas
e da crescente dependência do meio digital, recomendamos que as 
ações de segurança sejam priorizadas e as medidas tomadas revisadas
para a melhoria da segurança dos serviços providos pelos órgão públicos
 e, consequentemente, dos cidadãos.

\bibliographystyle{sbc}
\footnotesize
\bibliography{artigo}
\end{document}